# Job Placement Advisor Based on Turnaround Predictions for HPC Hybrid Clouds

Renato L. F. Cunha, Eduardo R. Rodrigues,
Leonardo P. Tizzei, Marco A. S. Netto
IBM Research

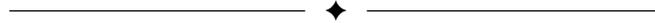

**Abstract**—Several companies and research institutes are moving their CPU-intensive applications to hybrid High Performance Computing (HPC) cloud environments. Such a shift depends on the creation of software systems that help users decide where a job should be placed considering execution time and queue wait time to access on-premise clusters. Relying blindly on turnaround prediction techniques will affect negatively response times inside HPC cloud environments. This paper introduces a tool to make job placement decisions in HPC hybrid cloud environments taking into account the inaccuracy of execution and waiting time predictions. We used job traces from real supercomputing centers to run our experiments, and compared the performance between environments using real speedup curves. We also extended a state-of-the-art machine learning based predictor to work with data from the cluster scheduler. Our main findings are: (i) depending on workload characteristics, there is a turning point where predictions should be disregarded in favor of a more conservative decision to minimize job turnaround times and (ii) scheduler data plays a key role in improving predictions generated with machine learning using job trace data—our experiments showed around 20% prediction accuracy improvements.

**Index Terms**—HPC, Cloud Computing, HPC Cloud, HPCaaS, Hybrid Cloud, Advisory System, Cloud Bursting, Machine Learning

## 1 INTRODUCTION

Cloud computing has become an essential platform for several applications and services, including those with High Performance Computing (HPC) requirements. A clear trend is the use of hybrid clouds comprising on-premise[1] and remote resources. During peak demands, jobs are submitted to the cloud rather than being submitted to on-premise clusters, which can have long queue waiting times compared to cloud resource provisioning times.

Current work on HPC cloud, also known as HPCaaS (HPC as a Service) [1], has mainly focused on understanding the cost-benefits of using cloud over on-premise clusters [2–8]. In addition, the aim has been to evaluate the performance gap between cloud and on-premise resources [9–12]. Even though cloud typically has slower internal network speeds than on-premise resources, bursting jobs to the cloud can still provide better overall performance in overloaded environments. Nevertheless, users still struggle to decide where to run their

1. In this work we use the terms "local" and "on-premises" interchangeably.

jobs at any given moment due to several factors. Supporting users in such a decision is the objective of this work.

Three factors are crucial for effective job placement in HPC cloud environments [13]: job queue waiting time, execution time, and the relative performance of the cloud compared to the performance of on-premises machines. If users knew how long their jobs would wait in the local queue and how long they would take to run in both environments, then they could obtain the optimal turnaround time. However, this information is not known in advance.

A common strategy is to estimate the waiting time and execution time using historical data. Nevertheless, those estimates are not always accurate. The existing prediction methods make mistakes and cannot be always trusted. In this paper, we propose an *advisor*: a tool that considers the uncertainty of the predictions to select which environment, either cloud or on-premises, the user should submit her jobs to. Based on a cloud *versus* on-premise performance ratio, this tool computes a turnaround time estimate for both environments and a measure of uncertainty. Only if the uncertainty is below a threshold, the user is advised to run in the environment with the shortest turnaround time, otherwise the advisor plays it safe by instructing the user to send the job to local resources.

The advisor processes historical logs of queueing systems to extract prediction labels, waiting and execution times, and features. Features can be either fields from the original logs, such as submission time, requested time and requested number of processors, or they can be derived from the queue state, e.g. queue size, processor occupancy, and queued work. Furthermore, scheduling promises are added to the mix of features and are shown to improve substantially the prediction accuracy.

The prediction is based on an Instance-Based Learning (IBL) algorithm [14] (a machine learning method) that relates a new incoming job with similar jobs in the history (data based predictions). The predicted waiting time, execution time and the uncertainty are computed by a function of the labels of similar jobs. These predictions are then combined in the advisor to make allocation decisions. We evaluated the advisor with traces of real job queues and show its benefits under different performance ratios having "saved-time" (§ 5.1) as the main evaluation metric.

The novelty of our work is a detailed study and resource management techniques to advance the state-of-the-art of HPC cloud. In summary, our main contributions are:

- Decision support tool based on runtime and wait time predictions for HPC hybrid cloud environments (§ 4);
- Cutoff function for conservative resource allocation decisions considering the uncertainty of execution and wait time predictions (§ 4, § 5);
- Evaluation of the advisor using traces from real supercomputing workloads with lessons learned on the management of prediction uncertainty. We also evaluated the advisor using real speedup curves from applications executed in six environments: three on-premises and three cloud-based (§ 5);
- Machine-learning enhanced predictor that exploits scheduling information as a feature (§ 5);
- Feature analysis impact on machine-learning-based predictions of job execution wait times (§ 5).

## 2 RELATED WORK

Research efforts related to our work are in three major areas: metascheduling, job waiting time prediction, and runtime prediction. Solutions from these areas come from cluster and grid computing but can be applied to HPC cloud.

### 2.1 Metascheduling

Literature on metascheduling is extensive, mainly from Grid computing and more recently on hybrid clouds and inter-cloud environments. For instance, De Assunção et al. [15] evaluated a set of scheduling strategies to offload jobs from on-premise clusters to the cloud. These strategies consider various backfilling approaches that check the job waiting queue's current status; decisions are made when jobs arrive and complete execution. Sabin et al. [16] studied metascheduling over a heterogeneous multi-site environment. Scheduling decisions rely on a job's expected completion time in each site; however errors on estimations of expected completion time are not considered in these decisions. Sotiriadis et al. [17] introduced a state-of-the-art review on metascheduling related technologies motivated by inter-cloud settings. The same research group presented a study on the role of meta-schedulers for inter-cloud inter-operability [18]. Garg et al. [19] introduced two heuristics for scheduling parallel applications in grids considering time and cost constraints. Malawski et al. [20] studied task planning over multiple cloud platforms using a mixed integer nonlinear programming having cost as their optimization goal.

### 2.2 Queue time predictions

Measuring how long a job will wait in a queue before its execution is a key component for deciding where jobs should be placed. There are several techniques available in the literature. For example, Li et al. [21] investigated methods and algorithms to improve queue wait time predictions. Their work assumes that similar jobs under similar resource states have similar waiting times as long as the scheduling policy and its configuration remains unchanged for a considerable amount of time.

Nurmi et al. [22] introduced an on-line method/system, known as QBETS, for predicting batch-queue delay. Their main motivation is that job wait times have variations that make it difficult for end-users to plan themselves and be productive. The method consists of three components: a percentile estimator, a change-point detector, and a clustering procedure. The clustering procedure identifies jobs of similar characteristics; the change-point detector determines periods of stationarity for the jobs; and the percentile estimator calculates a quantile that serves as a bound on future wait time.

Kumar and Vadhiyar [23] developed a technique that defines which jobs can be classified as *quick starters*. These are jobs with short waiting times compared to the other jobs waiting for resources. Their technique considers both job characteristics such as request size and estimated runtime, and the state of the system, including queue and processor occupancy states.

More recently, Murali and Vadhiyar [24] proposed a framework called Qespera for prediction of queue waiting times for HPC settings. The proposed framework is based on spatial clustering using history of job submissions and executions. The weights associated with the features for each prediction are adapted depending on the characteristics of the target and history jobs.

### 2.3 Runtime predictions

Smith [25] developed a method/system for estimating both queue wait time and job runtime predictions. The method is based on IBL techniques and leverages genetic algorithms (GA) to refine input parameters for the method to obtain more accurate predictions. This system is used by XSEDE[2] to predict queue wait time.

Yang et al. [26] proposed a technique to predict the execution time of jobs in multiple platforms. Their method is based on data collected from short executions of a job and the relative performance of each platform.

Tsafrir et al. [27] developed a technique for scheduling jobs based on system-generated job runtime estimates, instead of using user provided estimates. For the runtime estimates, they analyzed several workloads from supercomputer centers and found out that users tend to submit similar jobs over a short period of time. Therefore, their estimations are based on the average time of the previous two actual job runtime values.

We built on top of existing efforts in the literature to help users make decisions of job placement, via the proposed cutoff function, considering the estimated uncertainty of job execution and wait time predictions. We also investigated in details the effects of features of the machine-learning predictor, how to extend it using scheduler information, and performed experiments with an extensive set of scenarios and workloads from real environments. The next section contains details of opportunities and problems posed by HPC hybrid cloud environments.

## 3 PROBLEM DESCRIPTION

Due to the heterogeneity of jobs in several supercomputing settings, mixing on-premise and cloud resources is a natural way to get the best of the two environments. In HPC hybrid clouds, users can experience fast interconnections in on-premise clusters[3] and quick access to resources in the cloud. Hybrid clouds are also cost-effective since it is possible to keep a certain amount of on-premise resources and rent cloud resources to meet minimal and peak demands, respectively.

---

2. XSEDE - https://www.xsede.org/

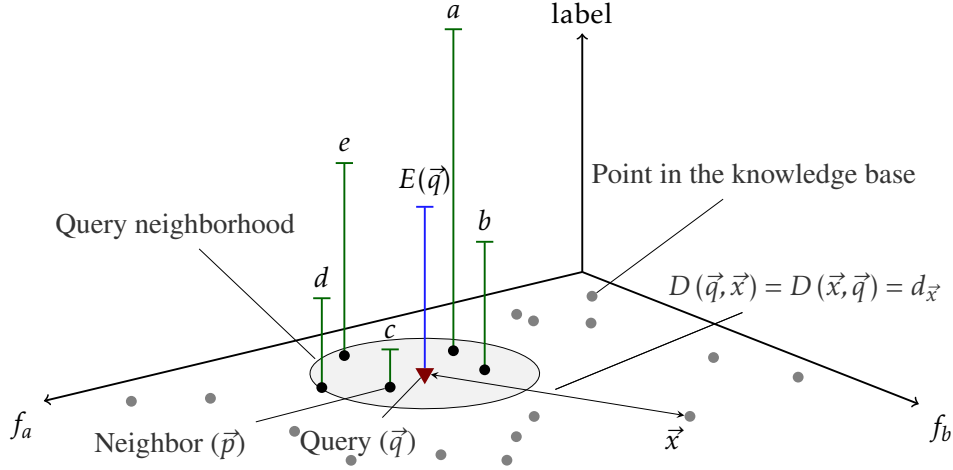

Fig. 1. Overview of the *k*-NN method used in this work. Axes $f_a$ and $f_b$ represent the features used, and the label axis represents the *values* recorded for those features (one notable exception is $E(\vec{q})$, which is the *predicted* (or *estimated*) value for the query, computed from the neighbors $a, b, c, d,$ and $e$). The points in black constitute the query point's neighborhood, while the points outside the neighborhood are represented in gray and without associated labels, to avoid polluting the image. Note that the radius of the neighborhood is not fixed, and will always be equal to the distance of the farthest neighbor of $\vec{q}$. In this case, this radius selects the $k = 5$ neighbors of the query point.

With this hybrid environment, one major challenge to users is to know where they should run their jobs. The decision involves several factors such as costs, raw performance on both environments, resource provisioning time in the cloud, and queue waiting time in the on-premise clusters. In this paper we envision a scenario where a company or supercomputing center has a contract with a cloud provider, which makes the cost to access cloud resources transparent to end-users, whose main concern is to optimize their jobs' turnaround time. The user is not wasteful, however, i.e. she never submits the same job to both environments. Nevertheless, the findings presented in this paper can be incorporated into other scenarios where users are totally exposed to cloud costs, or users receive quotas to use cloud resources; therefore they want also to minimize monetary costs.

The decision where to run a particular job would be simplified if execution times and cluster waiting times were known in advance and were accurate. In practice this is not possible since (i) job execution times depend on application input and underlying computing infrastructure, and (ii) waiting times depend on unpredictable scheduler queue dynamics created by other job arrivals and undetermined completion times.

The problem we tackle in this paper is therefore: *"how to predict job execution times and job waiting times, and more importantly, how to make job placement decisions based on the fact that these predictions can be inaccurate"*. We rely on state-of-the-art statistical models to make predictions, which are based on knowledge of past jobs and their characteristics.

3. Fast network interconnects, e.g. InfiniBand, are still not popular in public clouds.

Our scenario therefore consists of a company/institution with a set of resources controlled by a cluster management system (such as LSF, SLURM, TORQUE, or PBS) and a cloud provider. We explore scenarios with different performance ratios between both environments. The cluster management system contains records of already executed jobs. In our case we use traces from supercomputers available in the Parallel Workloads Archive [28]. Our work focuses mostly on CPU-intensive applications with no high data volumes to be transfered over the network—which is the case of popular applications in several industries such as EDA (Electronic Design Automation). For data-intensive jobs [29], we assume jobs have their data already in the cloud to be processed. For cases that additional time is required to transfer data to the cloud, the techniques proposed here can be further enhanced to include such a time. As a further note, prediction of transfer time is more accurate than runtime and waiting time according to Smith [25]. Consequently transfer time should only make the cloud an almost deterministic delta slower than the local environment.

## 4 ADVISOR TOOL: OVERVIEW AND METHOD

The focus of this work is on the development and evaluation of a decision-support tool, the *advisor*, for hybrid clouds built using prediction techniques from the literature [25]. These techniques are based on IBL [14] services that provide two types of predictions: wait time estimates (how long a job is expected to wait in a local resource manager queue before execution) and runtime estimates (how long a job is going to run, once it starts execution).

The advisor acts as a traditional resource management system: the user submits a job to the advisor, which calculates the best environment for job execution, and then forwards jobs to the actual resource manager. If the user disagrees with the advisor's suggestion, she can still force a job to execute in her chosen environment [13].

The next sections introduce the key ideas behind the estimators used in this work (§ 4.1), the technique for users

| Feature | Type | Description |
|---|---|---|
| User ID | Category | User who submitted the job |
| Group ID | Category | User group that submitted the job |
| Queue ID | Category | Number of the queue the job has been submitted to |
| Submission time | Number | Time at which the job was submitted |
| Requested time | Number | Amount of time requested to execute the job |
| Requested processors | Number | Number of processors requested at the submission time |
| Queue size | Number | Number of jobs in the wait queue at job submission time |
| Queued work | Number | Amount of work that was in the queue at job submission time |
| Remaining work | Number | Amount of work remaining to be executed at job submission time |
| Weekday | Number | Day of the week in which the job was submitted |
| Time since midnight | Number | Time of the day at which the job was submitted |
| Free processors | Number | Amount of free processors when the job was submitted |

to decide where to submit a job based on a cutoff function (§ 4.2), correctness issues we identified while implementing a state-of-the-art predictor from the literature (§ 4.3), and how we extended the predictor to provide uncertainty information which is required for the decision making process (§ 4.4).

## 4.1 Estimator design

In this section we briefly describe our implementation of the estimator found in the literature [25]. We used an algorithm known as $k$-Nearest Neighbors ($k$-NN) to generate the predictions. The principle behind this algorithm is, given a query, to find a predefined number of training samples closest in the feature space to the query and, then, to make a prediction using the $k$ nearest points found, as exemplified in Figure 1. Notice that $k$ is a hyper parameter of the model. In our experiments, it ranges from 6 to 33.

The learned examples are stored in a knowledge base, created in a training phase of the algorithm. The data stored in the knowledge base has two parts: a set of *input features* and a set of *output features* or *labels*. A feature is a variable that has a name and a type, and holds a value. Types usually are strings (nominal features), integers or floating-point numbers, and a feature's value belongs to the feature's type. Input features describe the conditions under which an event happened, and the output features describe what happened, in a dimension of interest, under those conditions. The features used by this algorithm for our job placement problem are described in Table 1.

To compute distances we used the Heterogeneous Euclidean-Overlap Metric [30], shown in Equation (1). This metric is a weighted sum over the distances of the various features, where the weights of each individual feature are represented by $w_{f_n}$ and the distance of each feature by $\delta_f$, which computes the ordinary euclidean distance for continuous variables and the overlap function[4] for categorical values. Each weight $w_{f_n}$ is learned by the system (§ 5.4). In the remainder of this text we will abbreviate the distance from query point $\vec{q}$ to point $\vec{x}$ as $d_{\vec{x}}$.

$$D(\vec{q}, \vec{x}) = d_{\vec{x}} = \sqrt{\sum_{f=1}^{N} \left( w_{f_n} \cdot \delta_f(q_f, x_f) \right)^2} \tag{1}$$

The computed distances are then passed to the kernel function, defined in Equation (2). Its purpose is to transform distances in a way that, as they approach zero, their transformed values approach a maximum constant value. Analogously, as distances increase, their transformed values approach zero exponentially fast. The kernel function has a parameter, denoted by $\omega$, that defines the kernel width and, therefore, how fast transformed distances will approach zero.

$$K(d_{\vec{x}}) = e^{-\left(\frac{d_{\vec{x}}}{\omega}\right)^2} \tag{2}$$

The actual prediction for a query point $\vec{q}$ is obtained by computing a weighted average of labels of the $k$ points closest to $\vec{q}$, where the weights are given by the kernel function from Equation (2). The complete estimation function is shown in Equation (3), where $l(\vec{p})$ is the label value of point $\vec{p}$, and the points $\vec{p}$ and $\vec{r}$ are the points from the set of $k$ nearest neighbors of $\vec{q}$. As also shown in Equation (3), each term $K(d_{\vec{p}}) / \sum_{\vec{r}} K(d_{\vec{r}})$ that multiplies $l(\vec{p})$ can be seen as a new weight in a weighted sum and can be further simplified as $v_{\vec{p}}$.

$$E(\vec{q}) = \frac{\sum_{\vec{p}} K(d_{\vec{p}}) \, l(\vec{p})}{\sum_{\vec{r}} K(d_{\vec{r}})} = \sum_{\vec{p}} v_{\vec{p}} l(\vec{p}) \tag{3}$$

## 4.2 Job placement decision based on a cutoff function

Figure 2 shows a sequence diagram of the major steps involved in job submission. Once the advisor receives a new job submission request $j$, it extracts features $\vec{f}$ from it, and predicts the time $j$ would be expected to wait ($w_p$) should it be scheduled to run in the on-premise cluster, and the expected time it would take for $j$ to run ($r_p$) in the same environment[5]. The output of each prediction is a pair *expected time, associated uncertainty*. The uncertainty is represented with a variance measure, and is denoted as $\sigma_w^2$ for the wait time predictor, and as $\sigma_r^2$ for the runtime predictor. As soon as the user makes her decision (either by accepting the advisor's suggestion or by forcing her preference), the job is submitted to the execution environment.

The decide function, shown in Figure 2, takes the outputs of the runtime ($r_p$ and $\sigma_r^2$) and wait time estimators ($w_p$ and $\sigma_w^2$) and decides whether a job should run locally or on the cloud[6]. Intuitively, it tests whether the predicted total time for executing a job in the local cluster ($w_p + r_p$) exceeds the time of running the same job in the cloud ($r^{cloud}(j, r_p)$). Since some predictions will be wrong, they should only be taken into account whenever the *estimated uncertainty* ($\sigma_w^2$) is smaller than a certain *cutoff* parameter (C($s$)), which depends on a parameter ($s$) that encodes the relative speed between environments. More formally, it implements Equation (4), where $r^{cloud}$ represents the estimated runtime in the cloud based on the characteristics of job $j$, such as the number of processors used, or amount of communication (§ 5.1.1 and § 5.1.2 present two possible models for $r^{cloud}$), and on the *predicted* run time for said job in the local environment. If the wait time variance is too high ($\sigma_w^2 \geq$ C($s$)), the advisor chooses to run $j$ locally.

---

4. The overlap function returns 1 when values are different and 0 otherwise.

5. The ordering of these calls is irrelevant, and these calls can be made in parallel, as processing only continues *after* both predictions are made.

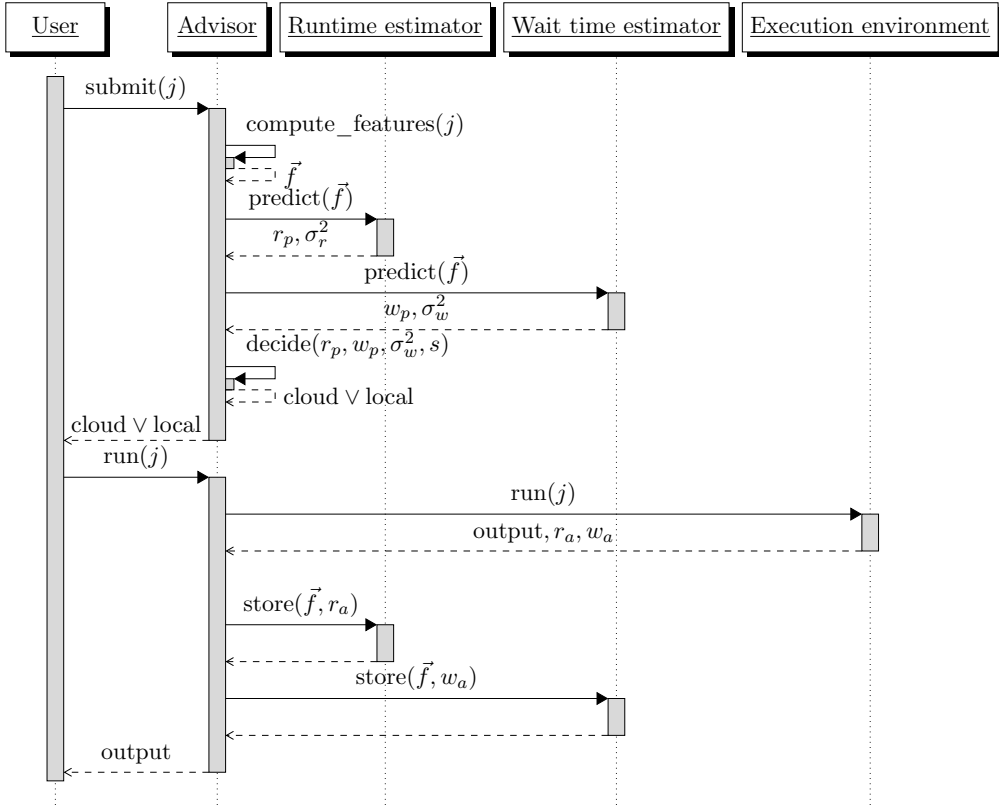

Fig. 2. Sequence diagram of operations performed by the advisor when a user submits a job. After predicting the run and wait times for a new job, the advisor presents its decision to the user for validation. The user, then, selects the execution environment, and the advisor submits the job for execution, storing actual run and wait times in the estimators as soon as the job is finished.

$$\text{decide}(r_p, w_p, \sigma_w^2, s) =$$
$$\begin{cases} \text{cloud} & \text{if } (w_p + r_p > r^{cloud}(j, r_p)) \wedge (\sigma_w^2 < \text{C}(s)) \\ \text{local} & \text{otherwise} \end{cases}$$

The cutoff parameter C($s$) is dynamic and depends on how fast an application running in the cloud compares to when run in on-premise resources. We empirically identified a power law dependency between speed and uncertainty tolerance. The reason for this observation is that applications that run much slower in the cloud must be carefully chosen to run off-premise, since any prediction mistake will have a very large impact in the overall performance. On the other hand, applications that run in the cloud close to local speed can tolerate larger mistakes. As a consequence, the advisor favors the local environment when estimates are too uncertain and the application runs slower in the cloud. We fitted an exponential function of the relative speed $s$, as displayed in Equation (4), with parameters $\alpha$ and $\beta$ fitted from a training set.

$$\text{C}(s) = \alpha e^{\beta s} \qquad (4)$$

In the training set there are fast and slow jobs (running in the cloud) with various variance values. Each job in the training set can be labeled either as a hit or a miss, depending on whether the advisor made a correct or incorrect prediction. We can then fit a linear boundary to the training set pairs ($s$, $\log(\sigma^2)$) that best separates hits and misses. The resulting line is given by the function $\log(\text{C}(s)) = \beta s + \log(\alpha)$. The fitted exponential serves as a boundary for predicting hits and misses. We subsequently used these parameters in the testing phase for evaluation. In this work, we decided to only consider the wait time variance for the cutoff. As will be discussed later, estimating the wait times of jobs is harder than estimating their runtimes and, thus, this variance accounts for most of the uncertainty of predictions.

### 4.3 Correctness issues when using floating-point arithmetic

If the estimator equations are implemented naïvely as found in the literature [25] and as described in Equations (2)–(3), the estimator will suffer from a major correctness issue: it has a non-zero probability of making divisions by zero. This derives from the fact that any number smaller than `FLOAT_MIN`, the smallest representable number in a floating-point system, is represented as zero [31]. Hence, the kernel function (2) will return zero whenever

$$d/\omega > \sqrt{-\ln(\texttt{FLOAT\_MIN})}. \qquad (5)$$

In an IEEE-754 environment with double precision (normalized and without gradual underflow), `FLOAT_MIN` $\approx 2.23 \times 10^{-308}$. So the kernel will be zero when $d/\omega \gtrapprox 26.62$.

A direct consequence from the result above is that whenever a query is sufficiently far from all its neighbors, the estimate weights $v_p$ will result in division by zero. For example, when $d$ is larger than 1 for all neighbors and $\omega < 1/\sqrt{-\ln(\texttt{FLOAT\_MIN})} \approx 0.038$.

Instead of treating this as a special case, there exists a more elegant fix: one can see that the individual weights $v_{\vec{p}}$ in Equation (3) can be rewritten as in Equation (6), which can never result in division by zero.

$$v_{\vec{p}} = \frac{K(d_{\vec{p}})}{\sum_{\vec{r}} K(d_{\vec{r}})} = \frac{e^{-(d_{\vec{p}}/k)^2}}{\sum_{\vec{r}} e^{-(d_{\vec{r}}/k)^2}} = \frac{1}{1 + \sum_{\vec{r}\neq\vec{p}} e^{(d_{\vec{r}}/k)^2 - (d_{\vec{p}}/k)^2}} \qquad (6)$$

### 4.4 Extending the estimator to provide uncertainty information

The $k$-NN method used finds the closest points to a query point in the feature space and computes a regression on the labels found in the knowledge base to output a prediction. Due to that, one is able to compute the *spread* of the data used for making a prediction. Intuitively, when the spread of the data is smaller, one can assign more confidence to the predictor's output. Analogously, when the spread of the data is larger, the confidence in the predictor is smaller. As such, for each prediction we output the spread of the data used for making the prediction, computed by the unbiased estimate of the data's variance, given as

$$s^2(\vec{q}) = \left( \frac{\sum_{\vec{p}} v_{\vec{p}}^2}{1 - \sum_{\vec{p}} v_{\vec{p}}^2} \right) \sum_{\vec{p}} v_{\vec{p}} (E(\vec{q}) - l(\vec{p}))^2 \qquad (7)$$

where $l(\vec{p})$ is the label of point $\vec{p}$.

## 5 EVALUATION

In this section we evaluate the advisor's effectiveness in helping a user to place jobs in HPC hybrid clouds. First we define the execution environment, workloads, and metrics used along with a theoretical model of the relative speed between the cloud and local environment, what we call speed ratio. Furthermore we define real speed ratios obtained from the literature of HPC cloud. Then we perform a series of analyses: (i) how the cutoff handles inaccurate predictions; (ii) the impact of using scheduling information to improve predictions; (iii) a comparison analysis between prediction speed-up and quality; and (iv) impact analysis of feature selection.

### 5.1 Environment setup, workload and metrics

We used workload logs from the Parallel Workloads Archive [28]; in particular the logs of San Diego Supercomputer Center's (SDSC) Blue Horizon (BLUE) and DataStar (DS) clusters, and from the High-Performance Computing Center North's Seth cluster (HPC2N). All these logs contain approximately $870k$ jobs in total. For all these workloads we used $500$ log entries for GA optimization. We also discarded the first $10k$ entries of the logs to prevent any cold-start effects and, once the estimators were trained, we evaluated them using $10k$ more log entries. These logs were used in two ways: (i) directly for training and evaluating the prediction services the advisor depends on and (ii) for replaying them in a scheduler simulator for testing the impact of scheduling promises on prediction quality (§ 5.3). Jobs moved to the cloud are not removed from the logs, so the job placement decision does not impact queue waiting time predictions. We focused then on the decision making at the submission time of each job.

In order to evaluate the advisor we need a cost function. One possibility is the Root Mean Squared Error (RMSE). Although one could argue that, intuitively, as we removed less certain predictions, the RMSE would drop, optimizing for this metric has a trivial solution: to make the cutoff so low that no job goes to the cloud. Consequently the RMSE would be made equal to zero, and an "optimal" solution would be achieved. Hence, this metric does not represent the advisor's real objective, that is to improve performance (reduce turnaround time) by means of running some jobs in the cloud.

Another possible metric to use could be accounting for hits and misses. With this metric, we would account whether predictions were confirmed or refuted by the real waiting time and execution time. Whenever the advisor told the user to send a job to the cloud based on the waiting time and execution time predictions and the actual execution time in the cloud was smaller than the local turnaround time, this would count as a hit, otherwise it would be a miss. The same procedure would be applied when the advisor suggested the user to go to the on-premise resources. Such a metric may be misleading though, since the advisor may perform well with small gain jobs, i.e. jobs whose saved time is small, and perform badly with jobs that cause large losses, giving the illusion of good performance without that being the case.

A better metric, and the one we settled on and called "saved-time", is to evaluate the time saved when submitting jobs to the cloud. If the advisor suggests the user to send a job to the cloud, we evaluate the time it takes for the job to execute in the cloud and subtract this value from the actual waiting time plus the actual execution time. Our base case for comparison is the *always-local* strategy, which represents a conservative user that only submits jobs to the local cluster. Therefore, if the advisor decides to run a job locally, the time it saved is equal to zero. The saved-time metric, thus, is

defined as:

$$T(j) = \begin{cases} 2\left((r^{\text{local}} + w_a) - r^{\text{cloud}}\right) & \text{if job } j \text{ runs in the cloud,} \\ 0 & \text{otherwise} \end{cases}$$

(8)

where $r^{\text{cloud}}$ is the run time of job $j$ in the cloud, $r^{\text{local}}$ is the run time of job $j$ in the local environment, and $w_a$ is the wait time in the local environment. $T$ can be negative, which represents a loss compared to the conservative always-local strategy.

Aside from the predictions, the advisor needs to evaluate the relative performance speeds of on-premise and cloud resources. The execution time in the on-premise resources can readily be obtained from the logs and corresponds to the *Run Time* field in the Standard Workload Format [28]. As for the cloud, the execution time depends on many factors. Differences in network interconnects play a dominant role in differences in performance, since clouds usually do not employ HPC-optimized networks. From this, it follows that as jobs with larger numbers of processors are submitted to the cloud, their performance deviate more from that seen in on-premise clusters. To account for these differences, we evaluate the advisor using two models described in the following sections: a linear theoretical model and an empirical model extracted from real speedup curves.

### 5.1.1 The linear model

The linear approach models the performance of the cloud as a linear function of the number of requested processors:

$$r^{\text{cloud}} = (o \cdot j_c + i) \cdot r^{\text{local}}$$

(9)

where $j_c$ is the number of used processors, $o$ is the overhead factor for each additional processor, and $i$ is the runtime component that is independent of the number of processors. The runtime component $i$ can incorporate the provisioning time and other factors that make the individual performance of cloud processors different from the local processor. In our experiments we considered that provisioning time is negligible compared to waiting time and runtime and the cloud processor is as fast as the local one. Consequently, $i = 1$ in our tests. Even though this is a simple model, it can be adjusted as the user submits jobs to the cloud. For our evaluations, this model corresponds also to the model of the reality, i.e. we do not adjust this model as applications are submitted to the cloud. The reason for this simplification is that we can isolate the evaluation of the advisor from the cloud model, which is a learning procedure in itself.

### 5.1.2 The empirical model

Our objective with using an empirical model is to have a better approximation of the advisor's performance with more realistic speed ratios between the cloud and local environments. To build this model, we extracted performance ratios from the work of Gupta et al. [32] in the form

$$s(j_c) = \frac{r_a^{\text{cloud}}(j_c)}{r_a^{\text{local}}(j_c)}$$

(10)

where $r_a^{\text{cloud}}$ and $r_a^{\text{local}}$ are the runtimes of a given application on cloud and local environments, respectively. Notice that in equation (10) we use the subscript $a$ to indicate the *actual* runtime, as opposed to the $p$ subscript, which is used to indicate *predicted* values.

In that work, the authors compared the runtime of 8 scientific applications on 3 cloud and 3 local environments. Hence, for each triple (cloud, local, application), we have a different $s(j_c)$, totalling $8 \cdot 3 \cdot 3 = 72$ performance ratios. Since that work only measured runtimes using number of processors ($j_c$) in the range $[1, 256]$, we limited our analysis to jobs that used a number of processors that fell in that range.

With these *ratios*, the model for predicting turnaround time is defined as

$$r_p^{\text{cloud}} = s(j_c) \cdot r_p^{\text{local}}$$

(11)

where $r_p^{\text{cloud}}$ and $r_p^{\text{local}}$ are the predicted runtimes of cloud and local environments respectively.

### 5.2 Cutoff: handling inaccurate predictions

The prediction of on-premise turnaround time depends on the prediction of both waiting time and execution time. Figure 3 contains heatmaps of actual and predicted runtime and waiting time, and we observe that the former due to highly-varied queue dynamics. Consequently, waiting time plays a more significant role in the uncertainty of the advisor's decision, and this is the reason for using a cutoff function that only takes the waiting time variance into account.

### 5.2.1 Analysis using theoretical speed ratios

The advisor's decisions for different overhead factors are presented in Tables 2, 3, and 4. For each log, we analyzed the time saved without the cutoff in the variance and with an exponential cutoff as a function of the overhead factor. The reason for doing so is that as the cloud becomes slower compared to the local environment, the user tolerates less uncertainty, because decision errors would waste more time. However, for more certain predictions the user might still want to submit jobs to the cloud. We compare the results with an optimal advisor (shown in the last three columns), which corresponds to the behavior of the advisor if it knew in advance the actual waiting and execution times of all jobs. Notice that even with high factors, some jobs can still be sent to the cloud, as shown in the "# of jobs on cloud" column of the optimal allocation.

There are some key differences among the analyzed systems. The jobs in the DS cluster are heavily penalized by slower clouds, whereas jobs from the HPC2N are not. In addition, when using the BLUE workload, the advisor presents its worst performance in the middle range of factors. These facts reflect the different characteristics of the jobs and how these differences influence the advisor. For example, more than 80% of jobs of DS are short jobs (less than 1 hour), even though only 30% have requested times that are less than 1 hour. While in HPC2N there are longer jobs and requested times are more similar to actual runtimes. These facts help the predictor and consequently the results are better for HPC2N. As for the BLUE workload, decisions are only affected when the overhead factor is sufficiently high, influencing the advisor to reduce the number of jobs sent to the cloud. In the middle range of factors, a sufficiently large

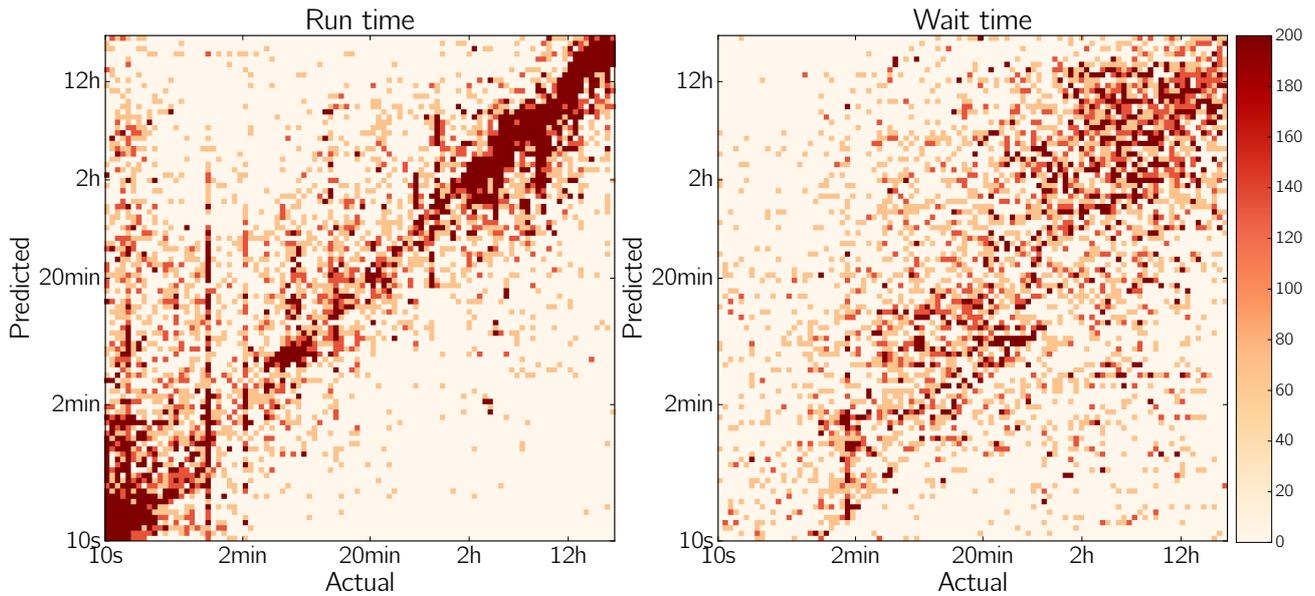

Fig. 3. Log log scatter plots of runtime predictions and wait time predictions for the HPC2N workload. Darker areas are more densely populated. Lighter areas are less densely populated. As can be inferred, runtime estimations are more accurate than wait time estimations.

TABLE 2
Job allocation decisions and saved time for the SDSC BLUE cluster.

| Factor | Without cutoff function | | | With cutoff function | | | Optimal allocation | | |
|---|---|---|---|---|---|---|---|---|---|
| | # of jobs on-premise | # of jobs on cloud | Saved-time (MM of sec.) | # of jobs on-premise | # of jobs on cloud | Saved-time (MM of sec.) | # of jobs on-premise | # of jobs on cloud | Saved-time (MM of sec.) |
| 0.01 | 4227 | 4318 | 130.05 | 4250 | 4295 | 123.96 | 4758 | 3787 | 155.98 |
| 0.05 | 6675 | 1870 | 12.54 | 7049 | 1496 | 2.21 | 6437 | 2108 | 80.34 |
| 0.10 | 7383 | 1162 | -1.86 | 7776 | 769 | 0.39 | 7118 | 1427 | 55.21 |
| 0.20 | 7904 | 641 | -10.71 | 8421 | 124 | 0.01 | 7607 | 938 | 40.86 |
| 0.30 | 8083 | 462 | -12.02 | 8534 | 11 | 0.00 | 7791 | 754 | 35.57 |
| 0.40 | 8219 | 326 | -4.02 | 8545 | 0 | 0.00 | 7929 | 616 | 32.49 |
| 0.50 | 8272 | 273 | -2.18 | 8545 | 0 | 0.00 | 8002 | 543 | 30.46 |
| 0.75 | 8362 | 183 | -3.04 | 8545 | 0 | 0.00 | 8100 | 445 | 26.99 |
| 1.00 | 8407 | 138 | -3.97 | 8545 | 0 | 0.00 | 8173 | 372 | 24.72 |
| 10.00 | 8531 | 14 | -0.40 | 8545 | 0 | 0.00 | 8446 | 99 | 10.21 |

TABLE 3
Job allocation decisions and saved time for the SDSC DS cluster.

| Factor | Without cutoff function | | | With cutoff function | | | Optimal allocation | | |
|---|---|---|---|---|---|---|---|---|---|
| | # of jobs on-premise | # of jobs on cloud | Saved-time (MM of sec.) | # of jobs on-premise | # of jobs on cloud | Saved-time (MM of sec.) | # of jobs on-premise | # of jobs on cloud | Saved-time (MM of sec.) |
| 0.01 | 1859 | 4692 | 131.95 | 1868 | 4683 | 130.95 | 2162 | 4389 | 177.75 |
| 0.05 | 3746 | 2805 | 27.44 | 4182 | 2369 | 11.69 | 3809 | 2742 | 134.95 |
| 0.10 | 4316 | 2235 | -37.27 | 5282 | 1269 | 0.87 | 4393 | 2158 | 110.72 |
| 0.20 | 4773 | 1778 | -135.92 | 5910 | 641 | -4.51 | 4921 | 1630 | 86.88 |
| 0.30 | 5092 | 1459 | -223.93 | 6307 | 244 | -3.53 | 5211 | 1340 | 74.38 |
| 0.40 | 5314 | 1237 | -266.48 | 6515 | 36 | -4.71 | 5442 | 1109 | 66.42 |
| 0.50 | 5444 | 1107 | -339.40 | 6549 | 2 | -0.01 | 5573 | 978 | 61.11 |
| 0.75 | 5714 | 837 | -435.65 | 6551 | 0 | 0.00 | 5804 | 747 | 54.14 |
| 1.00 | 5847 | 704 | -551.68 | 6551 | 0 | 0.00 | 5902 | 649 | 50.26 |
| 10.00 | 6341 | 210 | -1394.15 | 6551 | 0 | 0.00 | 6289 | 262 | 24.97 |

number of jobs is still sent to the cloud, which accounts for the losses seen in the table.

### 5.2.2 Analysis using real speed ratios

Next, we used the advisor to select the environment where to run jobs and, as in the previous section, we computed the time saved. Figure 4 shows saved-time as a function of the scenario, where each scenario corresponds to a triple (cloud, local, application)—notice, however, that there is no particular ordering between scenarios. The graph in the left side displays the results without the cutoff function, whereas the graph on the right shows time saved with an exponential cutoff function of the number of processors used.

TABLE 4
Job allocation decisions and saved time for the HPC2N cluster.

| Factor | Without cutoff function | | | With cutoff function | | | Optimal allocation | | |
|---|---|---|---|---|---|---|---|---|---|
| | # of jobs on-premise | # of jobs on cloud | Saved-time (MM of sec.) | # of jobs on-premise | # of jobs on cloud | Saved-time (MM of sec.) | # of jobs on-premise | # of jobs on cloud | Saved-time (MM of sec.) |
| 0.01 | 2659 | 7319 | 360.70 | 2659 | 7319 | 360.70 | 3497 | 6481 | 402.86 |
| 0.05 | 4612 | 5366 | 276.48 | 4634 | 5344 | 270.81 | 4928 | 5050 | 362.47 |
| 0.10 | 5669 | 4309 | 200.13 | 6313 | 3665 | 107.68 | 5649 | 4329 | 328.30 |
| 0.20 | 6617 | 3361 | 134.86 | 8406 | 1572 | 28.46 | 6391 | 3587 | 287.36 |
| 0.30 | 7170 | 2808 | 87.45 | 9210 | 768 | 11.37 | 6841 | 3137 | 262.54 |
| 0.40 | 7547 | 2431 | 61.50 | 9737 | 241 | 3.74 | 7119 | 2859 | 244.13 |
| 0.50 | 7810 | 2168 | 42.07 | 9797 | 181 | 5.74 | 7318 | 2660 | 229.89 |
| 0.75 | 8231 | 1747 | 12.23 | 9837 | 141 | 3.32 | 7692 | 2286 | 205.51 |
| 1.00 | 8459 | 1519 | 3.88 | 9851 | 127 | 4.99 | 7941 | 2037 | 189.75 |
| 10.00 | 9579 | 399 | -49.85 | 9933 | 45 | 0.13 | 9105 | 873 | 83.65 |

Without the cutoff function, decisions for the DS workload wasted time for many scenarios, whereas in the BLUE and HPC2N workloads, time was saved for almost all scenarios. In order to clarify what caused those results we plotted the average speed ratio between cloud and local for each one of the scenarios (Figure 5). The average speed ratio used in HPC2N is much smaller than in the others, since most of the jobs in HPC2N used few processors. Consequently, the ratios are small and the penalty for eventual allocation errors does not affect much the time saved metric. This, however, does not explain the difference between DS and BLUE. For these two workloads, the reason for the difference lies on the fact that the predictions for BLUE overestimate the wait time of the jobs. Consequently, more jobs are sent to the local environment. As a result, the advisor becomes more conservative and its behavior is closer to the always-local strategy. In the DS workload, predictions do not have a clear pattern, and the advisor's mistakes outweighs the times in which it makes correct decisions, resulting in unsatisfactory performance. Nonetheless, with the addition of the cutoff function, the advisor is able to decide with greater certainty when to run jobs in the cloud, as one can see in the right side of Figure 4. This reflects the fact that, for high ratio values, the advisor takes a conservative approach and becomes similar to the always-local strategy.

## 5.3 Exploiting scheduling information to improve predictions

As can be gathered from analyzing Figure 3, there is some room for improvement in the quality of the wait time predictions. One way of doing so is by integrating *scheduling* information into the prediction model. Some scheduling algorithms, such as the conservative backfilling algorithm [33], maintain an upper bound (henceforth called "promise") of when a given job is going to start its execution. Therefore, promises generated at the time of job submission can be added as a feature for the predictor to use.

As we do not have the scheduler details of the workloads we used, we processed them with an in-house discrete-event simulator. From the simulation, we extracted the promises generated by a conservative backfilling algorithm and added those promises as a feature to the estimator. The effects of such change to the estimator can be seen in Table 5. Although using promises alone (column 2) is worse than using the baseline estimator (column 1), integrating the promises into the estimator actually make it better (column 3).

TABLE 5
Evaluation of different wait time estimators. The first column shows the estimator's behavior without modifications. The second column illustrates how it would fare in case it predicted the exact same values of the scheduler's promises. The results in the third column were generated by integrating the promises as a new feature in the predictor.

| Workload | Root Mean Squared Error (s) | | |
|---|---|---|---|
| | Baseline | Promises | Promises as a feature |
| SDSC BLUE | 12223.96 | 24859.58 | 9828.45 |
| SDSC DS | 9345.64 | 20812.19 | 7598.18 |
| HPC2N | 8346.05 | 8471.45 | 6911.02 |

A direct consequence of enriching the estimator with promises is that the advisor can now make better decisions, as observed in Figure 6: the advisor, when using promises as a feature, is consistently better than the baseline advisor. The only case in which this does not happen is when the performance factor equals zero, and the reason is simple: since that model considers that cloud resources can be provisioned in no time, any estimate of queue wait time would make the advisor prefer the cloud. When promises were added, wait time predictions for that case went down, making the advisor prefer the local environment, resulting in "losses" in saved time for the advisor that used promises. With the other factors, the improvement in the quality of the predictions was enough to make the advisor perform better.

## 5.4 Speeding-up predictions

Feature weights, number of neighbors, kernel width, and knowledge base size can be adjusted to better fit the labels (either runtime or waiting time). Ideally, these parameters should be adjusted on-line, i.e. as new jobs start or finish to run, they should be included in the runtime or waiting time training set respectively, and new parameters should be computed.

Finding optimal parameters is not straightforward though. The objective function is highly nonlinear and discontinuous. For example, changing the feature weights may change the neighbors of a given query dramatically, completely changing the value of the prediction. In our previous experiments, we generated the parameters by using a GA approach. Murali and Vadhiyar [24] use a computationally cheaper approach: they compute the correlation between each feature and labels and use this value as their estimator's weights. In this section, we compare this strategy with the GA optimization approach, considering both how much faster models can be trained and how accurate such models are.

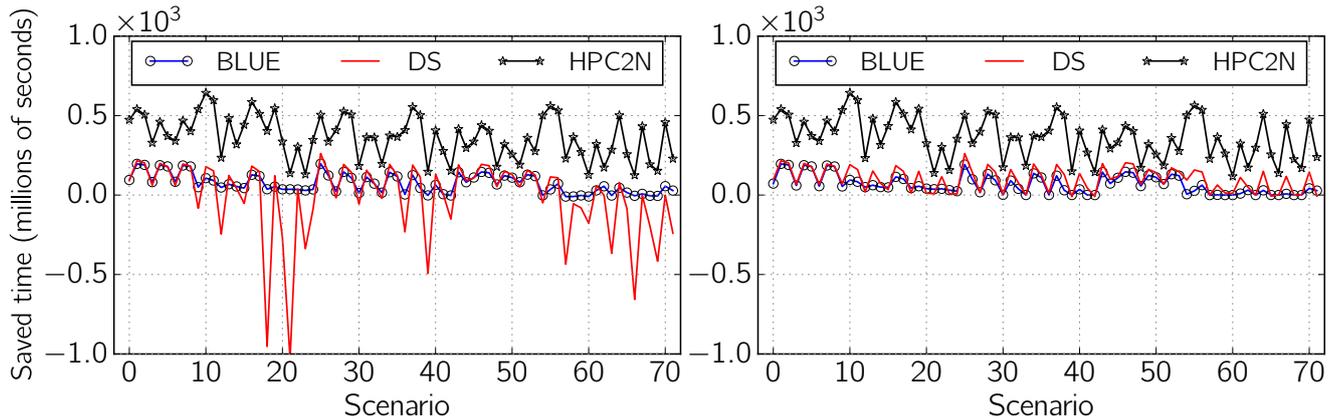

Fig. 4. Saved time for all 72 scenarios with and without the cutoff function.

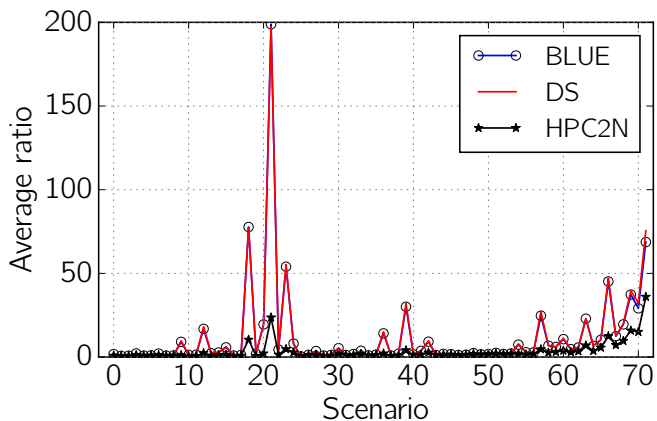

Fig. 5. Average ratio as a function of scenario.

A GA chromosome is defined as $[k, \omega, s, w_{f_1}, \ldots, w_{f_n}]$, where $k$ is the number of neighbors, $\omega$ is the width parameter of the kernel function, $s$ is the size of the knowledge base, and $w_{f_1}$ to $w_{f_n}$ are the weights of features $f_1$ to $f_n$. For scoring, the RMSE function was used. The GA was run with a population of size 50 for 100 generations. The probability of crossover was 80% and the mutation probability was 10%. Elitism was set to 5%. The running time for the optimization process to finish was approximately 10 hours in a current hardware environment.

The correlation approach for computing weights uses the intuitive idea that features that are highly correlated to the label should have a large weight in the regression. This approach may not capture the non-linear behavior of the fit, but it has the major advantage of being much faster than GA. Fitting the same weights as before runs in only 0.15 seconds. Table 6 shows the accuracy comparison between

GA and the correlation approach for the logs we have used. Apart from the wait time of HPC2N, the correlation results are very similar to the ones obtained with GA. One possible explanation for these results is that the simpler correlation captures most of the variability of the label and consequently the less biased GA does not improve much the results.

## 5.5 Feature selection analysis

In this section we discuss the impact of feature selection in the quality of the run time and wait time predictors. To do so, we performed *best subset selection*, a technique for finding the best model for a given learning task. In this setting, we consider models that have a subset of the features described in Table 1, plus the scheduler promise, described previously (§ 5.3), and *user jobs*, a feature that counts how many jobs of the submitting user are currently in the system. Therefore, for all $k \in \{1, \ldots, 14\}$, we evaluated all the $\binom{p}{k}$ possible models, with $p = 14$, the number of features from Table 1 plus the promise and user jobs features. The models in question used the correlation method described in the previous section for finding the weight vector, and were evaluated using the RMSE metric.

The results are summarized in Figure 7 where, in the top row, we see the performance of the run time predictors and, at the bottom, we see the performance of the wait time predictors. As can be concluded from analyzing the figure, each workload has a different optimal number of features which, somewhat counter-intuitively, does *not* equal the performance of using all 14 features. This can be explained by noticing that, since weights are computed based on pairwise correlation between the feature values and the labels, the model ends up assigning non-zero weights to features that affect the model's performance negatively. When performing *best subset selection*, the model ends up removing this negative interaction, improving its performance.

The features, along with their weights, used in the models that exhibited the best performance are shown in Table 7. As can be seen in the table, the scheduler promise is present in all of the models with best performance for wait time prediction, reinforcing the notion that this is, indeed, an important feature. Since the HPC2N cluster has only a single queue, the *queue id* is absent from the best model for that workload. Also notice that, for the HPC2N workload, promises are so good that the best model learned uses them as sole feature. This happens due to the fact that the vast majority of jobs require a small number of processors and a small amount of time. That is, most jobs are similar and, therefore, scheduler promises are, indeed, quite similar to reality, making them the key feature for predictions.

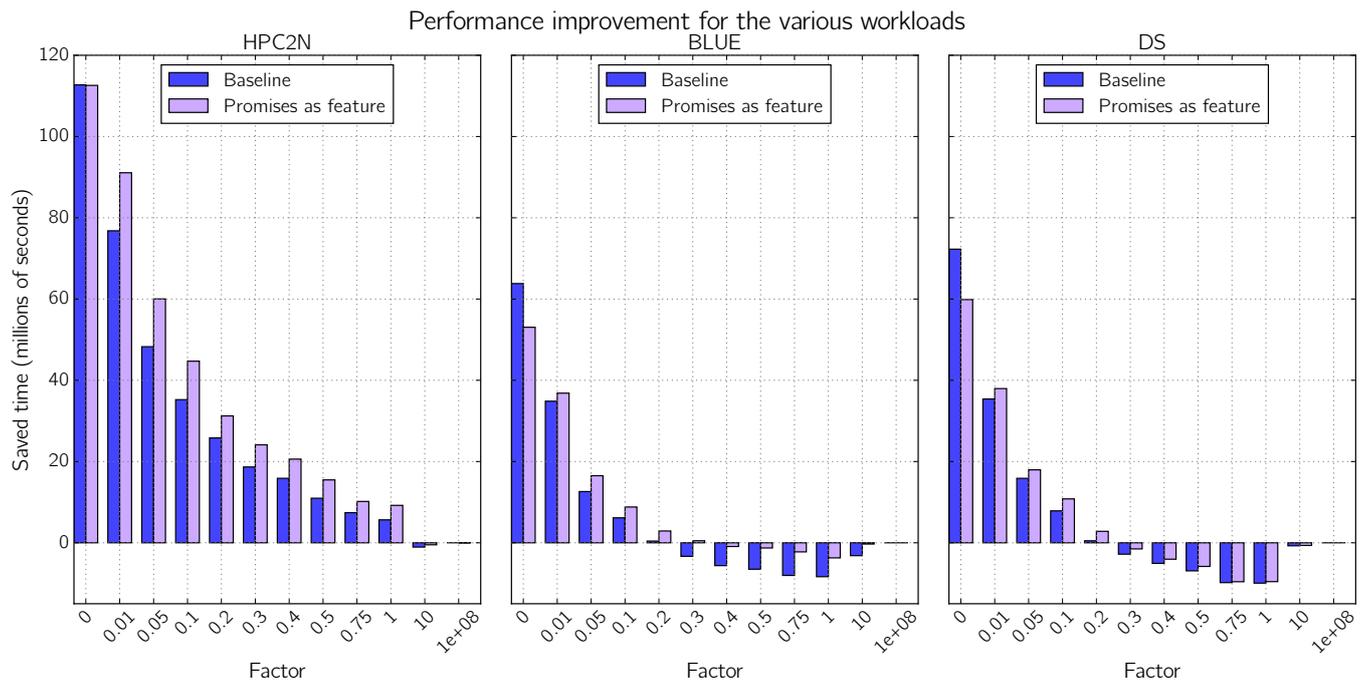

Fig. 6. Improvement of the advisor's performance for each factor of the linear model for each workload. "Baseline" refers to the advisor's decisions when using the default wait time predictor. "Promises as feature" refers to the advisor's decisions when using the scheduler's promises as a feature.

TABLE 6
Accuracy comparison between GA and correlation.

| Label | Workload | Correlation | GA | Improvement over GA (%) |
|---|---|---|---|---|
| Run time | DS | 16226 | 16074 | 0.94 |
| | BLUE | 12222 | 12169 | 0.43 |
| | HPC2N | 36295 | 36403 | -0.29 |
| Wait time | DS | 97259 | 97109 | 0.15 |
| | BLUE | 70223 | 70239 | -0.02 |
| | HPC2N | 84579 | 74800 | 13.07 |

TABLE 7
Features used by the model with best performance in each workload.

| Label | Workload | Features | Weights |
|---|---|---|---|
| Run time | BLUE | user ID, requested time, queued work, remaining work, time since midnight | 0.01, 0.4, 0.12, 0.35, 0.09 |
| | DS | user ID, group ID, requested processors, requested time, promise, queued work, remaining work | 0.19, 0.14, 0.07, 0.21, 0.02, 0.07, 0.14 |
| | HPC2N | submission time, requested processors, requested time | 0.07, 0.29, 0.26 |
| Wait time | BLUE | queue ID, requested processors, promise | 0.44, 0.57, 0.98 |
| | DS | queue ID, weekday, promise, queue size, user jobs, queued work, remaining work | 0.07, 0.12, 1, 0.69, 0.79, 0.79, 0.05 |
| | HPC2N | promise | 1 |

## 5.6 Threats to validity

Threats to external validity are conditions that limit our ability to generalize the results of our experiment to industrial practice [34]. We identified two main threats to external validity. The first one is that workload logs used in this study might not be representative of scientific HPC environments. We minimized this threat by selecting three distinct workload logs (i.e., BLUE, DS and HPC2N), which were collected from real clusters and contain hundreds of thousands jobs. Furthermore, our analysis has shown key differences among the analyzed systems (§ 5.2.1), which also minimizes this threat.

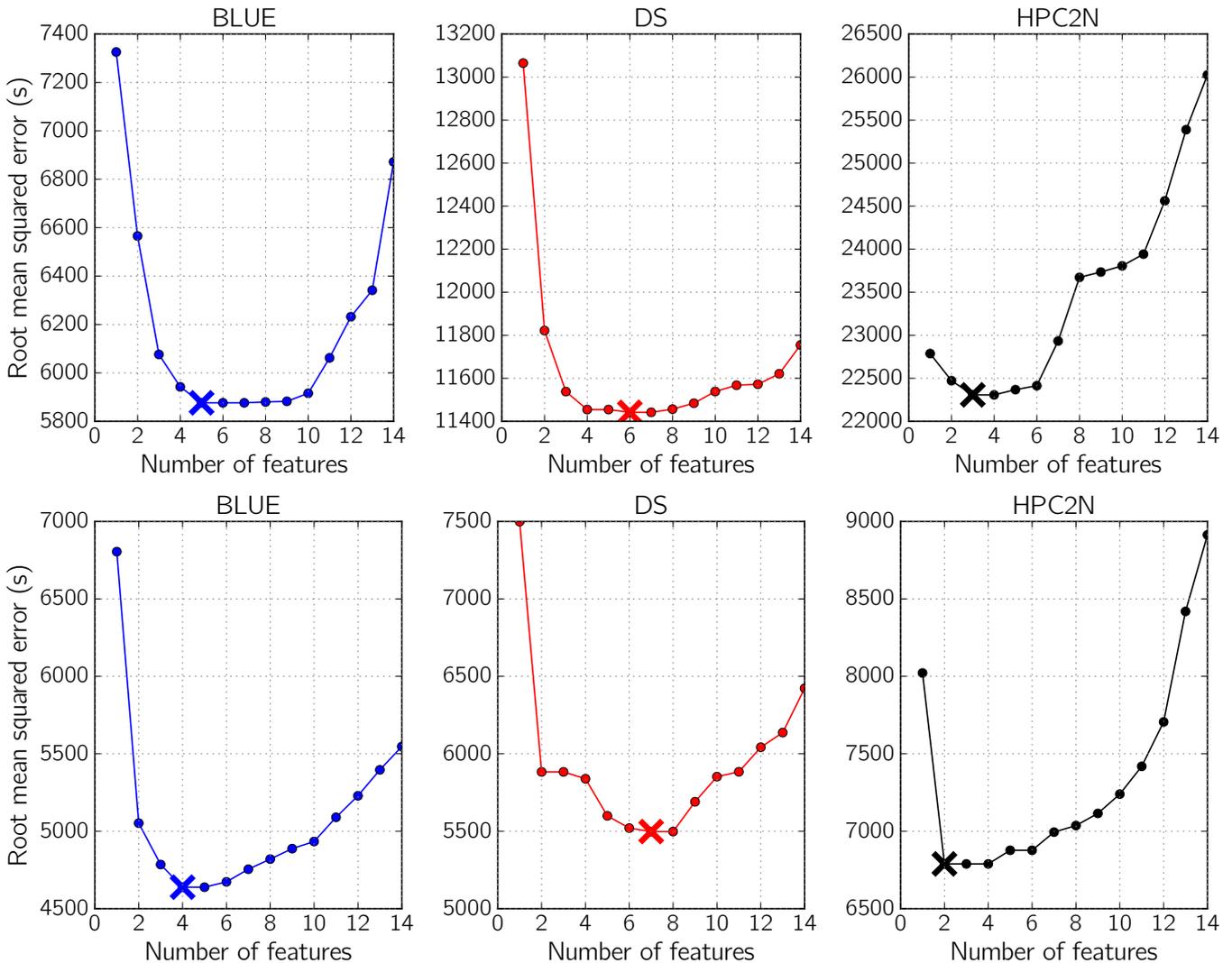

Fig. 7. Root Mean Squared Error for the best models of each size for each of the workloads for run time predictions (top) and wait time predictions (bottom). The cross denotes the method with best performance. Where there was a draw, models with less features should be preferred.

Since it is not feasible to determine the runtime on cloud environment of all the jobs that were collected from workload logs, we relied on speed ratios to estimate such runtime. These speed ratios might not be representative of ratios between real clusters and cloud providers, thus threatening external validity of this study. We minimized this threat by building an empirical model that uses all possible combinations of 3 cloud and 3 cluster environments that executed 8 different scientific applications (§ 5.1.2). These combinations generated a wide spectrum of speed ratios that were used in this study.

### 5.7 On using the advisor in real systems

One of the key components in the advisor is the variance cutoff (§ 4.2). The other key component is the model of the performance ratio between the local on-premise cluster and the cloud resources (§ 5.1.1–5.1.2). None of which may be available when a system based on the advisor is just deployed, as these models use need data to be fitted.

The most straightforward way of building such a model would be by doing comprehensive benchmarks of applications in the cloud, which might not be cost-effective. This problem can be solved incrementally by submitting a few pilot jobs of the most commonly executed applications to the cloud and recording execution statistics such as the ones used as features in this work. Once the advisor is put into operation with this "crude" model, it can be further updated and refitted as more jobs are submitted to the cloud by the advisor.

## 6 CONCLUSION

We proposed an advisor to help users decide where to run jobs in HPC hybrid cloud environments. The decision relies on queue waiting time and execution time of the jobs, which are predicted using traces from past job scheduling data. Even though we have used state-of-art predictors, the estimations of wait time and runtime can be inaccurate.

One of our contributions is to define a cutoff function of the uncertainty, limiting the impact of prediction errors when using cloud resources. This cutoff establishes the maximum uncertainty tolerated by the user to accept predicted results. Beyond this value the advisor takes a conservative approach of submitting a job to on-premise resources. The function is parametrized by a speed factor. The logic behind it is that the user tolerates less uncertainties when the cloud is slower and, consequently, the penalty for errors are larger.

Some lessons learned in this work are: (1) the performance difference between cloud and on-premise resources has direct impact on the relevance of an advisory system to help users make job placement decisions. Very imbalanced environments (i.e. high performance for either cloud or on-premise clusters) may simplify the decision choice of the users but can have considerable losses in turnaround time if wrong decisions are made. Therefore, traps can be minimized for imbalanced environments and assistance can be given to users in more difficult decisions when the environments have similar performance by using an advisory tool as the one proposed in this paper. (2) in an imbalanced environment, the cutoff function must take into account the relative speeds of both the cloud and local environments. We empirically found out that this relationship follows an exponential decay, i.e. as the cloud environment gets slower than the on-premise cluster, the cutoff value must decay exponentially with the decrease in relative speed. Caused by the mistakes made by the predictor.

From a prediction perspective, we can draw the following lessons: (1) simply increasing the knowledge base's size with more past data may decrease the performance of the predictor, this is due to the loss of temporal locality and, hence, past data actually behaves like noise, hurting prediction quality; and (2) floating-point precision plays an important role in prediction services and, therefore, must be carefully implemented. Some of these issues may not appear until the service is in production.

Finally, from an industry perspective, we are observing that the demand for hybrid clouds is growing at a fast rate. The work presented in this paper can be effectively used by the HPC community, even if cloud resources are not as powerful as on-premise clusters. The work presented here aims to help users to make better resource allocation decisions, but we believe there is still room for improvements. In particular, better predictors can improve the confidence of users on where to place their jobs and consequently increase the demand for this type of HPC hybrid cloud services.

## ACKNOWLEDGEMENTS

We thank Warren Smith for providing anonymized data used to validate our estimator implementation. We thank Carlos Cardonha, Matthias Kormaksson, Sam Sanjabi for reviewing early drafts of this work, and the anonymous reviewers. We also would like to thank Dror G. Feitelson for maintaining the Parallel Workload Archive, and all organizations and researchers who made their workload logs available. This work has been partially supported by FINEP/MCTI under grant no. 03.14.0062.00.